\documentclass[prb,twocolumn]{revtex4}

\usepackage{graphicx}

\renewcommand\hat\widehat

\begin{document}

\title{Thermodynamics of itinerant magnets in a classical spin fluctuation model}

\author{A. L. Wysocki}
\affiliation{Department of Physics
and Astronomy and Nebraska Center for Materials and Nanoscience,
University of Nebraska--Lincoln, Lincoln, Nebraska 68588, USA}
\author{J. K. Glasbrenner}
\affiliation{Department of Physics and Astronomy and Nebraska Center
for Materials and Nanoscience, University of Nebraska--Lincoln,
Lincoln, Nebraska 68588, USA}
\author{K. D. Belashchenko}
\affiliation{Department of Physics and Astronomy and Nebraska Center
for Materials and Nanoscience, University of Nebraska--Lincoln,
Lincoln, Nebraska 68588, USA}

\date{\today}

\begin{abstract}
Thermodynamics of itinerant magnets is studied using a classical model with one
parameter characterizing the degree of itinerancy. Monte Carlo simulations for
bcc and fcc lattices are compared with the mean-field approximation and with
the Onsager cavity field approximation extended to itinerant systems. The
qualitative features of thermodynamics are similar to the known results of the
functional integral method. It is found that magnetic short-range order is weak
and almost independent on the degree of itinerancy, and the mean-field
approximation describes the thermodynamics reasonably well. Ambiguity of the
phase space measure for classical models is emphasized. The Onsager cavity
field method is extended to itinerant systems, which involves the
renormalization of both the Weiss field and the on-site exchange interaction.
The predictions of this approximation are in excellent agreement with Monte
Carlo results.
\end{abstract}

\maketitle

\section{Introduction}

The thermodynamics of magnetic materials is often described using the
Heisenberg model in which the spins are attached to lattice sites. Real magnets
are much more complicated, because the magnetization is due to band electrons
whose degree of localization varies between different materials. This so-called
\emph{itinerancy} manifests itself in the fluctuation of the magnitudes of the
local moments, which may be defined in a muffin tin sphere or using a
projection in an appropriate basis. Thus, the degree of itinerancy may be
characterized by the relative importance of longitudinal and transverse
(rotational) fluctuations of the local moments. \cite{Moriya} In the localized
(Heisenberg) limit the longitudinal spin fluctuations (LSF) have a large energy
scale and are suppressed. This limit is approached in some magnetic insulators.
Metals, on the other hand, are often quite far from this limit, because the
exchange splitting and the bandwidth are typically of the same order.
Experimentally, itinerancy is most clearly revealed in the paramagnetic
susceptibility by the deviation of the effective moment found from the
Curie-Weiss constant from the true local moment, as well as by the deviations
from the Curie-Weiss law.

A large amount of work has been devoted to the thermodynamics of
itinerant magnets using phenomenological Ginzburg-Landau models for
weak ferromagnets \cite{MD,LT,Moriya} or the Hubbard model and the
functional integral methods. \cite{MT78,Hubbard,Hasegawa,Moriya}
These studies have clarified the role of LSF in thermodynamics and
explained the observed behavior of the paramagnetic susceptibility.
However, these methods are unsuitable for quantitative studies of
realistic materials. Ginzburg-Landau expansions, as is well known,
correctly describe only the contribution of long-wave fluctuations
and must always be rigged with a wavevector cut-off. Such models are
useful in the studies of critical phenomena, but they are irrelevant
to the determination of the critical temperature itself, which is
determined by \emph{short-range} fluctuations. \cite{LL5} An
unsatisfactory signature of Ginzburg-Landau models is the absence of
any information on the short-wave components of the exchange
interaction in the resulting expressions for the Curie
temperature.\cite{MD,LT,MW} In our opinion, the neglect of
short-wave fluctuations in these models makes their predictions for
magnetic short-range order (MSRO) also unreliable. The functional
integral method, on the other hand, suffers from the necessity to
make severe and ambiguous approximations.\cite{HE73}

Magnetic thermodynamics has also been studied using density functional theory
(DFT) by treating spin fluctuations within the adiabatic approximation
\cite{Gyorffy} assuming that the relevant fluctuations are well represented by
constrained \cite{Dederichs} noncollinear ground states. The most widespread
approach is the disordered local moment (DLM) approximation
\cite{Oguchi,Gyorffy} which relies on the single-site approximation and is
designed to approximate the DFT ground state of a system with random directions
of the local moments. The LSF have been neglected in all implementations of
this approach so far, restricting its application to magnets which are close to
the localized limit. In particular, the DLM method neglecting LSF fails for
(strongly itinerant) nickel where it finds vanishing local moment in the
paramagnetic phase.\cite{Staunton}

Other authors studied \emph{itinerant} thermodynamics by mapping the
results of first-principles energies for various spin configurations
(including both transverse and longitudinal fluctuations) to a
classical Hamiltonian in which variable local moments play the role
of dynamical variables, and then exploring the thermodynamics of
this Hamiltonian using either the variational principle in
reciprocal space \cite{Uhl} or Monte Carlo (MC) simulations in real
space. \cite{Rosengaard,Lezaic,Ruban} These calculations clearly
show that LSF, as expected, are very important in nickel. Moreover,
they revealed only weak magnetic short-range order (MSRO) above the
Curie temperature $T_c$ for both Fe and Ni, which is similar to the
Heisenberg model. These results are consistent with the fact that in
any lattice model with no frustration all correlation corrections to
the mean-field approximation (outside of the critical region) should
be small in the parameter $1/z$, where $z$ is the number of
neighbors within the interaction range. \cite{DMFT} On the other
hand, very strong MSRO above $T_c$ was found \cite{Antropov} in Ni
using the \emph{ab initio} spin dynamics method, which, similar to
DLM, is based on the adiabatic approximation and neglects LSF.

Classical models with variable local moments seem to capture the
important qualitative features of the thermodynamics of itinerant
magnets which are similar to the predictions of the functional
integral method. However, these models have been built and studied
only for a few particular materials, and a general study of their
thermodynamic properties has not been undertaken. Such a study is
useful as a step to more refined models with the advantage that
numerically exact results for a classical model are easily
accessible through Monte Carlo simulations. Therefore, in this paper
we explore the thermodynamics of a classical spin fluctuation model
as a function of the degree of itinerancy using MC simulations and
simple analytic approximations. We emphasize that here we are not
concerned with the ``mapping'' procedure (which can be quite
challenging) but rather focus on the other separate part of the
program, i.e. on the determination of thermodynamics once the
Hamiltonian has been defined. We therefore restrict ourselves to the
simplest possible realization of this model which includes only
\emph{one} free parameter characterizing the degree of itinerancy.

\section{Model}

Our model is a lattice version of the phenomenological model of Murata and
Doniach \cite{MD} written with a vector order parameter \cite{Moriya}:
\begin{eqnarray}
H=\frac12\sum_\mathbf{q}\chi^{-1}(\mathbf{q})\mathbf{m}_\mathbf{q}\mathbf{m}_{-\mathbf{q}}+\frac
B4 \sum_i
m_i^4\nonumber\\
=\sum_{i}\left[\frac12\left(\chi^{-1}_{00}-I\right)m^{2}_{i}+\frac B4
m_i^4\right]-\frac{1}{2}\sum_{i\ne j}J_{ij}\mathbf{m}_{i} \mathbf{m}_{j}.
\label{Hamiltonian}
\end{eqnarray}
Here $\mathbf{m}_{i}$ denotes the magnetic moment at site \emph{i}
whose length is unrestricted, and $I$ the Stoner
exchange-correlation parameter. We have separately written the local
$\chi^{-1}_{00}=\partial B_i/\partial m_i$ and nonlocal
$J_{ij}=-\chi^{-1}_{ij}$ parts of the unenhanced inverse
susceptibility. This model involves a number of simplifying
assumptions: (1) It is classical in the sense that $\mathbf{m}_i$
are dynamical variables and not operators. (2) Both local and
nonlocal parts of the inverse susceptibility are considered to be
independent of the magnetic state and \emph{isotropic}. In general,
$\chi^{-1}_{ij}$ is a Cartesian tensor which depends on the magnetic
state and reduces to a scalar only in the paramagnetic state. (3)
Nonlinear effects are included only through a local fourth-order
term, similar to the Murata-Doniach model.

Model (\ref{Hamiltonian}) is somewhat similar to that used to
represent the unified spin fluctuation theory\cite{MT78} classically
(see Ref. \onlinecite{Moriya}, Ch. 7, and also Ref.
\onlinecite{Kubler}), with an important difference: the energy of
LSFs is included as a function of local dynamical variables
$\mathbf{m}_i$, rather than that of one global parameter $\langle
m^2_i\rangle$. This difference is similar to that between the
Heisenberg model and the spherical approximation to it.

In the ground state all local moments are parallel and we recover
the Stoner model which is ferromagnetic if $IN(E_F)>1$, where
$N(E_F)=\chi(0)$ is the density of states at the Fermi level in the
nonmagnetic state. This Stoner criterion can also be written as
$(I+J_0)>\chi^{-1}_{00}$ where $J_0=\sum_j J_{ij}$. On the other
hand, in the paramagnetic or non-magnetic matrix, local moments
exist in the Anderson sense only if $I>\chi^{-1}_{00}$ which is
stricter than the Stoner criterion. We will call this the Anderson
criterion. (Note that $\chi^{-1}_{00}\ne1/\chi_{00}$.)

Introducing reduced local moments $\mathbf{x}_{i}=\mathbf{m}_i/m_0$, where
$m_{0}$ is the value of all $m_i$ at $T=0$, the Hamiltonian (\ref{Hamiltonian})
can be conveniently parameterized:
\begin{equation}
H' \equiv \frac{H}{J_{0}m^{2}_{0}} = \sum_{i}E(x_{i}) - \frac{1}{2}\sum_{i\neq
j}\frac{J_{ij}}{J_0}\mathbf{x}_{i}\cdot \mathbf{x}_{j} \label{ReducedH}
\end{equation}
where $E(x)=[ax^2/2+bx^4/4]/J_0$ with $a=\chi^{-1}_{00}-I$ and
$b=Bm_0^2=J_0-a$. For the nearest neighbor model with coordination
number $z$ we have $J_{nn}/J_0=1/z$, and for the given lattice $H'$
contains only one parameter, which we define as $\alpha = \arctan
b/a$. Note that $b>0$ is equivalent to the Stoner criterion, and
$a<0$ is equivalent to the Anderson criterion. \cite{Antropov-pc}

To understand the meaning of the parameter $\alpha$, consider the
ground state of Hamiltonian $H$ with a single-site excitation, i.e.
the state with $\mathbf{m}_i=\mathbf{m}_0$ for all $i$ except $i=c$.
The energy of this state has a minimum at
$\mathbf{m}_c=\mathbf{m}_0$ and its curvature with respect to the
longitudinal fluctuation of $\mathbf{m}_c$ is $K_\parallel=J_0+2b$,
while the curvature with respect to transverse fluctuations is
$K_\perp=J_0$. Their ratio $K_\parallel/K_\perp=1+(2b/J_0)$
characterizes the relative importance of longitudinal and transverse
fluctuations. If $b\gg J_0$, the fluctuations are mainly transverse,
and we have the localized (Heisenberg) limit for which $a\approx -b$
and $\alpha\approx3\pi/4$. If $b\ll J_0$, the transverse and
longitudinal spin fluctuations are equally important; this limit
corresponds to $\alpha=0$. The Anderson criterion is equivalent to
$\alpha>\pi/2$. Thus, the parameter $\alpha$ characterizes the
degree of itinerancy and is similar to those appearing in other
theories \cite{MT78,Moriya}. Note that we always have
$K_\parallel/K_\perp>1$, even though the \emph{macroscopic}
longitudinal stiffness is proportional to $b$ and tends to zero at
$\alpha\to0$.

Evaluation of the thermodynamic properties involves taking a trace over the
quantum states, or a functional integral over the classical degrees of freedom.
To our knowledge, in all classical models reported so far and based on \emph{ab
initio} calculations, the uniform measure in the space of $\mathbf{m}_i$ was
used. \cite{Uhl,Rosengaard,Lezaic,Ruban} However, our dynamical variables are
not canonical, and therefore the phase space measure (PSM) is not known. In the
case when LSF are important, the PSM has to be supplied along with the
Hamiltonian as an additional phenomenological ingredient. Strictly speaking, it
is not possible to disentangle the measure from quantum statistics; for
example, in the atomic limit only integer moments with atomic multiplet
degeneracies should be present. Ambiguity of PSM is intrinsic to all
microscopic classical spin fluctuation models including the classical version
of the ``unified theory'' of Moriya and Takahashi (Ref. \onlinecite{Moriya},
Sec. 7) and its extensions, \cite{Kubler} as well as the functional integral
approach combined with the static approximation which destroys the correct
quantum operator properties. In the latter case, the Hubbard-Stratonovich
transformation can be applied with the interaction term written in different
ways, which produce different results after the static approximation is made.
\cite{Hubbard,HE73} Two particular choices discussed by Hubbard \cite{Hubbard}
result in different measures in the space of fluctuating fields $\mathbf{v}_i$:
uniform in one case, and involving the weighting factor $\prod_i v_i^{-2}$ in
another. To explore the influence of PSM on thermodynamics, we will consider
these two measures in the space of the local moments $\mathbf{m}_i$.

\section{Thermodynamic properties: Monte Carlo and mean-field results}

Monte Carlo simulations for model (\ref{ReducedH}) were performed
using the Metropolis algorithm for bcc and fcc lattices with nearest
neighbor exchange. At each step the new random direction and
magnitude of the moment on one site was tried, and sampling of the
moment magnitude was performed according to the chosen PSM. We used
supercells with up to $3456$ or $6912$ sites for bcc or fcc lattices
($12\times12\times12$ unit cells with periodic boundary conditions).
The reduced Curie temperature $t_c=T_c/(J_0m_0^2)$ was found using
the fourth-order cumulant method \cite{Binder}, and the paramagnetic
susceptibility was calculated using the fluctuation-dissipation
theorem.

In the mean-field approximation (MFA) the magnetization is found from the
self-consistency condition $\langle x_z\rangle =\partial \ln Z_1/\partial
(\beta h_W)$, where

\begin{equation}
Z_1=\int_{0}^{\infty}g(x)x\frac{2\sinh(\beta h_W x)}{\beta h_W}e^{-\beta
E(x)}dx
\end{equation}

is the single-site partition sum, $h_W=\langle x_z\rangle$ the reduced Weiss
field, and $g(x)$ the weighting factor, which is either 1 or $x^{-2}$ for the
two chosen PSM's. $E(x)$ is defined after Eq. (\ref{ReducedH}), and $\beta=1/t$
is the inverse reduced temperature.

\begin{figure}[htb]
\begin{center}
\includegraphics[width=0.45\textwidth]{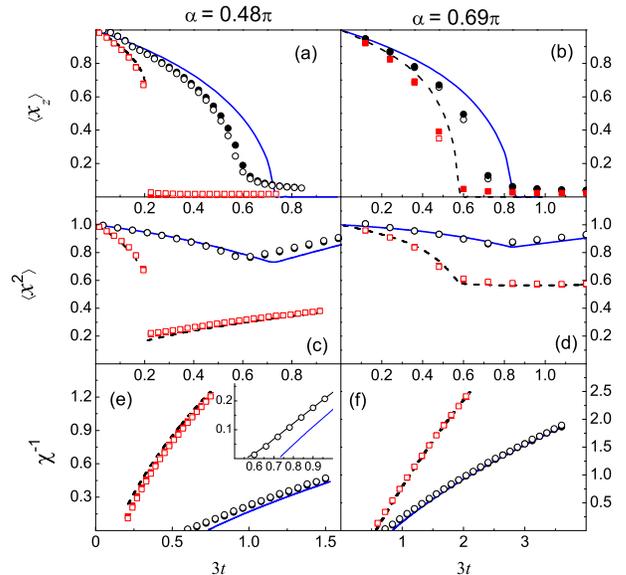}
\end{center}
\caption{(Color online) (a-b) Reduced magnetization $\langle x_z\rangle$, (c-d)
mean squared local moment $\langle x^2\rangle$, and (e-f) inverse paramagnetic
susceptibility $\chi^{-1}$ as a function of the reduced temperature
$t=T/(J_0m_0^2)$. MFA results are shown by solid (blue online) lines for
$g(x)=1$ and by dashed black lines for $g(x)=x^{-2}$. MC results are displayed
by black circles for $g(x)=1$ and by red (gray) squares for $g(x)=x^{-2}$ (in
both cases the symbols are filled for fcc and empty for bcc lattice). The inset
in panel (e) highlights the region close to $t_c$ for the bcc lattice with
$g(x)=1$ and also shows the results of the generalized Onsager method (black
line connecting the MC points).} \label{fig1}
\end{figure}

Fig. \ref{fig1} shows the temperature dependence of magnetization,
the average square of the local moment and the paramagnetic
susceptibility using the reduced variables according to Eq.
(\ref{ReducedH}). Results are shown for two values of $\alpha$:
$0.48\pi$ and $0.69\pi$. In both cases the agreement between MC and
MFA results is very good for all properties (MFA overestimates $T_c$
by 20\% or less). The results strongly depend on PSM, especially in
the more itinerant case $\alpha=0.48\pi$. In particular, for the
uniform PSM a second-order phase transition occurs for both values
of $\alpha$, but for the PSM with $g(x)=x^{-2}$ the phase transition
is of first order for $\alpha=0.48\pi$, and $T_c$ is nearly 2.8
times smaller compared to that for $g(x)=1$.

As seen in Fig. \ref{fig1}, below $T_c$ the average $\langle x^2\rangle$
declines with temperature due to the decrease of the Weiss field, which causes
the maximum of the distribution function to shift to smaller moments. This is
in agreement with earlier results
\cite{Moriya,Hubbard,Hasegawa,Uhl,Rosengaard,Ruban}. The width of the
distribution function increases with temperature, which counteracts the
decrease of the local moment. The PSM with $g(x)=x^{-2}$ puts less weight on
the states with large moments, and hence $\langle x^2\rangle$ drops much faster
compared to the uniform PSM. If the Anderson criterion is not satisfied
($\alpha<\pi/2$) then the most probable moment in the paramagnetic state is
zero. In this case, $\langle x^2\rangle$ increases with temperature above $T_c$
as seen in Fig. \ref{fig1}c. On the other hand, if the Anderson criterion is
satisfied, the local moment may slightly decrease in a range of temperatures
above $T_c$, as seen for $g(x)=x^{-2}$ in Fig. \ref{fig1}d.

The magnetic susceptibility above $T_c$ is shown in Figs. 1e,f. In MC
simulations it is calculated using fluctuation-dissipation theorem, while in
MFA we directly consider the response of the system to the external magnetic
field. Excellent agreement between MFA and MC is observed except for the small
error in $T_c$. In MFA one obtains above $T_c$
\begin{equation}
\chi_{\mathrm{MFA}} = \frac{\frac{1}{3}\langle x^2
\rangle}{t-\frac{1}{3}\langle x^2 \rangle} \label{susceptibility}
\end{equation}
This formula looks similar to the Curie-Weiss expression in the Heisenberg
model, but here $\langle x^2\rangle$ depends on temperature, which leads to a
renormalization of the CW constant and deviations from the CW law. The CW
constant $C=d\chi^{-1}/dt$ (for a second-order phase transition) is now given
by
\begin{equation}
C = \frac{3}{\langle x^2(t_c) \rangle}\left[1 - \left.\frac{d\log\langle x^2
\rangle}{d\log t}\right|_{t_c}\right]\label{Curieconstant}
\end{equation}
Thus, in addition to the usual Heisenberg term the Curie constant has a
contribution due to the temperature dependence of $\langle x^2 \rangle$ (second
term in square brackets in (\ref{Curieconstant})). As a result, the effective
moment squared $x_\mathrm{eff}^2=3/C$ deviates from $\langle x^2\rangle$. As
discussed above, $\langle x^2\rangle$ usually increases with temperature above
$T_c$, which, according to Eq. (\ref{Curieconstant}), reduces $C$ and increases
$x_\mathrm{eff}^2$. Moreover, for the uniform PSM $\langle x^2\rangle$
increases faster with temperature compared to PSM with $g(x)=x^{-2}$, and hence
the CW constant is much smaller in this case (see Fig. 1f and also 1e, where
the transition is however of first order).

\begin{figure}
\begin{center}
\includegraphics[width=0.45\textwidth]{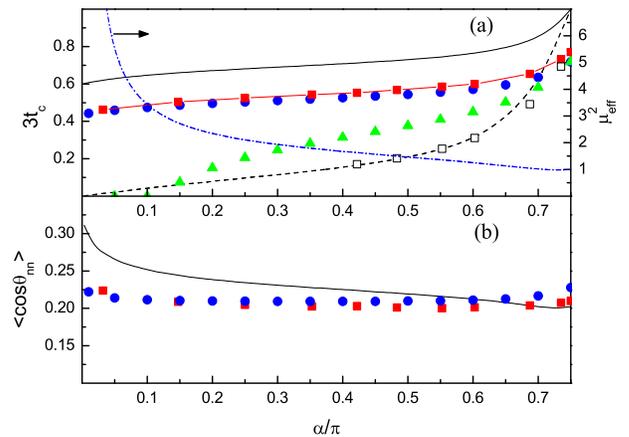}
\end{center}
\caption{(Color online) (a) Reduced Curie temperature $t_c$ and (b) MSRO
parameter $\langle\cos\theta_{nn}\rangle$ at $T=1.1T_c$ as a function of the
itinerancy parameter $\alpha$ for the bcc lattice. Solid black line, red (gray)
squares, and blue (dark gray) circles show the results of MFA, MC, and the
generalized Onsager method for $g(x)=1$, respectively. Dashed black line and
empty black squares depict MFA and MC results for $g(x)=x^{-2}$. Green (light
gray) triangles represent the incomplete Onsager reaction field correction with
the on-site interaction left unrenormalized. The blue (gray) dash-dotted line
in the upper panel shows the effective moment $x^2_\mathrm{eff}$ found from the
Curie constant for $g(x)=1$ in MFA. Very similar results were obtained for the
fcc lattice (not shown).}\label{fig2}
\end{figure}

In Fig. \ref{fig2} some thermodynamic properties of the system are plotted as a
function of the itinerancy parameter $\alpha$. From Eq. (\ref{susceptibility})
it follows that the MFA value of $t_c$ for the second-order phase transision is
found by solving the equation $3t_c=\langle x^2(t_c)\rangle$, where $\langle
x^2(t)\rangle$ is fully determined by $E(x)$ in Eq. (\ref{ReducedH}). This is
an easy way to estimate $T_c$ for an itinerant system using first-principles
data for $E(x)$, $J_0m_0^2$, and the assumed PSM. However, for PSM with
$g(x)=x^{-2}$ the transition is of first order except for a small region close
to the local moment limit (in MFA the tricritical point where the order of the
phase transition changes is at $\alpha_{tr}=0.632\pi$). Therefore, in general
one must consider the minima of the free energy as a function of the
magnetization, which can also be easily done in MFA. Note that the order of the
phase transition depends on the details of the model and can change if, for
example, the dependence of the exchange parameter on the magnetization is taken
into account. In particular, the phase transition for the model of Ni is of
first order in Ref. \onlinecite{Uhl} (as seen from the abrupt drop of $M(T)$
and $M_s$ at $T_c$ in their Fig. 2) and in Ref. \onlinecite{Ruban} (as seen
from the abrupt drop of $\overline m$ in their Fig. 6), even though the uniform
PSM was used in both of these models.

From Fig. \ref{fig2} we see that when the transition is of second
order, MFA overestimates $T_c$ by about 20\%, which is typical for
the Heisenberg model. When the transition is of first order, MFA
gives an almost exact $T_c$. It is important that even for the
second-order transition the overestimation of $T_c$ in MFA does not
depend on the degree of itinerancy. This is consistent with the fact
that the degree of MSRO, which is shown in Fig. \ref{fig2}b for
$T=1.1T_c$, is quite small and stays essentially constant in the
whole range of $\alpha$. Thus, in our model itinerancy does not lead
to strong short-range order. This result agrees with Refs.
\onlinecite{Rosengaard,Ruban} where weak short-range order was found
for the models of Fe and Ni. Note that if the exchange interaction
extends to more than one shell of neighbors and stays mainly
ferromagnetic, the MFA validity criterion is satisfied even better,
and the MSRO parameter should further decrease. Similar to the
Heisenberg model, strong MSRO may only be expected in
low-coordinated lattices or in the presence of frustration when for
some pairs $J_{ij}/kT_c$ is not small.

The square of the effective moment $x^2_\mathrm{eff}$ is also shown
in Fig. \ref{fig2} for the uniform PSM (dash-dotted line). In the
local limit $x_\mathrm{eff}$ naturally tends to 1. However, as
$\alpha$ is decreased towards zero, the ratio
$x^2_\mathrm{eff}/\langle x^2(t_c)\rangle$ increases and eventually
becomes much larger than 1. Similar behavior is found in functional
integral theories. \cite{Moriya}

\section{Generalized Onsager correction for itinerant systems}

Onsager introduced the concept of a \emph{cavity field} in the theory of polar
liquids, which is designed to go beyond the molecular field approximation (MFA)
by including short-range order effects. \cite{Onsager} The cavity field is the
\emph{effective} internal field which orients polar molecules in the
ferroelectric phase. Onsager observed that each molecule polarizes the
surrounding liquid and thereby generates a \emph{reaction field} acting back on
the molecule. However, this field is always parallel to the molecule's dipole
moment and hence does not affect its orientation. Therefore, for a liquid with
\emph{permanent} dipoles the reaction field must be subtracted from the mean
molecular field, the result being the cavity field. Onsager also noted that the
reaction field enhances the dipole moments of real molecules due to their
polarizability.

The cavity field method was successfully applied to Ising
\cite{Brout} and Heisenberg \cite{Logan} magnets which have
permanent magnetic moments. Cyrot \cite{Cyrot} noted that
Moriya-Kawabata's self-consistent renormalization theory for the
Hubbard model may be essentially reproduced by using Onsager-like
arguments; more recently this method was implemented numerically.
\cite{Cyrot96} However, the actual physics there is very different;
Cyrot's approach seeks the correlation correction with respect to
the Hartree-Fock solution, which is unrelated to short-range order.
Onsager's method was also applied to itinerant nickel,
\cite{Staunton} but, as we will see below, correct generalization to
itinerant systems with LSF requires an additional ingredient which
was missed in Ref. \onlinecite{Staunton}.

We now generalize Onsager's method to magnets with LSF described by Hamiltonian
(\ref{Hamiltonian}). Consider model (\ref{Hamiltonian}) above $T_c$ in a small
external collinear magnetic field $H^{ext}_i\mathbf{e}_z$. We pick site $0$ and
integrate out the degrees of freedom from all the other sites in the partition
function to obtain the effective Hamiltonian in the form of a generating
functional for the lattice with a cavity \cite{DMFT}. Expanding this functional
around the atomic limit to order $1/z$ we obtain

\begin{eqnarray}
H^0_{\mathrm{eff}}=E(m_0)-\mathbf{m}_0\left(\mathbf{H}^{ext}_0+\sum_iJ_{0i}\langle\mathbf{m}_i^c\rangle\right)\nonumber\\
-\frac{m_0^2}{2}\sum_{ij}J_{0i}J_{0j}\chi_{ij}^c \label{Heff}
\end{eqnarray}
where the superscript $c$ refers to the lattice with a cavity, i.e.
with site $0$ removed, and we used the fluctuation-dissipation
theorem to express the pair correlator through the susceptibility.

In order to find the magnetization and susceptibility of the lattice with a
cavity we need to solve the ``impurity problem.'' Using the linked-cluster
expansion technique,\cite{Wortis} the longitudinal susceptibility of the
original lattice can be written as follows:
\begin{equation}
\hat\chi=\hat\Pi+\hat\Pi\hat W\hat\Pi \label{Tmat}
\end{equation}
where $\hat W$ is the effective interaction that satisfies the equation $\hat
W=\hat J+\hat J\hat\Pi \hat W$, and $\hat\Pi$ is the 1-bond-irreducible
``polarization operator'' which may be shown to be local to first order in
$1/z$. \cite{VLP} (All quantities in Eq. (\ref{Tmat}) are matrices in site
indices.) Removal of site $0$ may be formally represented by a perturbation
$\Delta\hat\Pi=-\Pi_{00}\delta_{0i}\delta_{0j}$ to $\hat\Pi$. (The
renormalization of $\Pi_{jj}$ for $j\ne0$ due to removal of site $0$ is at
least of order $1/z^2$.) Thus, denoting the effective interaction matrix for
the cavity lattice as $\hat W_c$, we may write $\hat W_c^{-1}-\hat
W^{-1}=-\Delta\hat\Pi$. Using (\ref{Tmat}) and the fact that $\hat\Pi$ is
diagonal, we find
\begin{equation}
\chi^c_{ij}=\chi_{ij}-\frac{\chi_{i0}\chi_{0j}}{\chi_{00}}. \label{chic}
\end{equation}
The average local moments $\mathbf{M}^c_i=\langle\mathbf{m}^c_i\rangle$ for the
lattice with a cavity are:
\begin{equation}
M_i^c=\sum_j\chi^{c}_{ij}H^{ext}_j=M_i-\frac{\chi_{i0}}{\chi_{00}}M_0,
\label{Sic}
\end{equation}
where $M_i$ are the average local moments of the complete lattice without the
cavity. The value of $\mathbf{H}^{ext}_0$
does not affect $M_i^c$ (as expected), therefore in the right-hand side of
(\ref{Sic}) we may take $M_i$ and $M_0$ for the actual field distribution.

From the effective Hamiltonian (\ref{Heff}) we can find the
magnetization at site $0$:
\begin{equation}
M_0=\tilde\chi^0\tilde H_W
\end{equation}
where
\begin{equation}
\tilde H_W=H^{ext}_0 +
\sum_iJ_{0i}\left(M_i-\frac{\chi_{i0}}{\chi_{00}}M_0\right)
\end{equation}
is the renormalized effective field (cavity field), and $\tilde\chi^0$ is the
\emph{renormalized} bare (atomic-limit) susceptibility. The latter may be
written as $\tilde\chi^0=\langle m^2 \rangle_\lambda/3T$, where the average
paramagnetic squared local moment $\langle m^2 \rangle_\lambda$ is calculated
using a renormalized on-site exchange $\widetilde{I}=I+\lambda$ with
$\lambda=\sum_{ij}J_{0i}J_{0j}\chi_{ij}^c$. This renormalization of the bare
susceptibility is the essential ingredient needed to extend Onsager's theory to
itinerant magnets. It has no effect in the localized limit where $m^2$ is
constant.

As usual, we now obtain the Fourier transform of the susceptibility:
\begin{equation}
\chi_\mathbf{q}=\frac{\tilde\chi^0}{1-\tilde\chi^0(J_\mathbf{q}-\lambda)}\;,
\label{Suscq}
\end{equation}
where
$\lambda=\sum_{\mathbf{q}}J_\mathbf{q}\chi_\mathbf{q}/\chi_{00}$. We
used the same symbol $\lambda$ as above in the definition of
$\widetilde I$, because these expressions are identical, as can now
be shown with the help of Eqs. (\ref{Suscq}) and (\ref{chic}). Eq.
(\ref{Suscq}) with the definitions of $\lambda$, $\tilde\chi^0$ and
$\widetilde I$ form a closed set of equations for the paramagnetic
susceptibility. Note that (\ref{Suscq}) \emph{automatically} leads
to a sum rule $\chi_{00}=\tilde\chi^0$, which agrees with the
fluctuation-dissipation theorem.

At the Curie temperature $\chi_\mathbf{q}$ diverges at
$\mathbf{q}=0$. Therefore, from (\ref{Suscq}) we obtain
$T_c=\frac{1}{3}J_0\langle m^2(T_c) \rangle_\lambda/G$, where
$G=\sum_{\mathbf{q}}(1-J_\mathbf{q}/J_0)^{-1}$ is the diagonal
element of the lattice Green's function. \cite{Logan} Note that the
value of $\lambda$ at $T_c$ is equal to $J_0(1-G^{-1})$ and
independent of the degree of itinerancy $\alpha$.

The reduced Curie temperature $t_c$ and MSRO parameter
$\langle\cos\theta_{nn}\rangle$ at $T=1.1T_c$ calculated in this way
are shown in Fig. \ref{fig2} for the bcc lattice and the PSM with
$g(x)=1$. The agreement with MC results is excellent in the whole
range of $\alpha$. We repeated these calculations for the fcc
lattice and found excellent agreement with MC as well. The accuracy
of the predicted $t_c$ may be seen from Table 1. Similar performance
for bcc and fcc lattices suggests that this approximation is not
very sensitive to the connectivity of the lattice. The paramagnetic
susceptibility is also shown in Fig. \ref{fig1}e for
$\alpha=0.48\pi$, bcc lattice, and uniform PSM. The agreement with
MC results is essentially perfect outside of the narrow critical
region.

\begin{table}[h]
\caption{Reduced Curie temperature $t_c$ for bcc and fcc lattices
for PSM with $g(x)=1$: Results of the mean-field approximation
(MFA), Horwitz-Callen approximation (HC), generalized Onsager method
(GO) and Monte Carlo (MC).}
\begin{tabular}{|c|c|c|c|c|c|c|c|c|}
\hline
$\alpha/\pi$ & \multicolumn{4}{|c|}{bcc} & \multicolumn{4}{|c|}{fcc}\\
\cline{2-9}
 & MFA & HC & GO & MC & MFA & HC & GO & MC\\
\hline 0.032 &0.621 &0.449 &0.451 &0.462(1)&0.621 &0.465 &0.466 &0.480(2) \\
\hline 0.148 &0.660 &0.484 &0.486 &0.504(2)&0.660 &0.501 &0.502 &0.520(5) \\
\hline 0.250 &0.681 &0.503 &0.504 &0.525(2)&0.681 &0.519 &0.520 &0.540(5) \\
\hline 0.352 &0.699 &0.518 &0.520 &0.543(2)&0.699 &0.535 &0.536 &0.562(2) \\
\hline 0.422 &0.712 &0.529 &0.530 &0.553(1)&0.712 &0.546 &0.547 &0.570(5) \\
\hline 0.483 &0.723 &0.539 &0.541 &0.568(1)&0.723 &0.557 &0.558 &0.584(2) \\
\hline 0.553 &0.745 &0.555 &0.557 &0.585(2)&0.745 &0.572 &0.574 &0.600(1) \\
\hline 0.602 &0.765 &0.570 &0.573 &0.600(2)&0.765 &0.589 &0.590 &0.617(2) \\
\hline 0.687 &0.834 &0.619 &0.622 &0.654(3)&0.834 &0.640 &0.642 &0.672(6) \\
\hline 0.735 &0.942 &0.683 &0.688 &0.732(2)&0.942 &0.708 &0.711 &0.753(6) \\
\hline 0.750 &1 &0.713 &0.718 &0.770\cite{MCheisenberg} &1 &0.740 &0.743 &0.788(3) \\
\hline
\end{tabular}
\end{table}

The first-order terms in the $1/z$ expansion derived above introduce two
corrections to MFA. The first one is the subtracted \emph{mean} reaction field;
this correction reduces the magnetization. This is the only correction in
Onsager's method for systems with permanent moments. The second correction
described by the last term in Eq. (\ref{Heff}) adds back the \emph{fluctuating}
reaction field which is always parallel to the moment on the central site. For
the Heisenberg (or Ising) model this second correction has no effect, but in
itinerant systems it always increases the local moments and hence the Curie
temperature. There is a strong cancelation between these two corrections in
itinerant systems, and improvement compared to MFA may be achieved only if both
of them are included. Indeed, if the renormalization of the Stoner parameter is
not taken into account (i.e. if the last term in Eq. (\ref{Heff}) is dropped),
we find a spurious strong suppression of $T_c$ for itinerant systems, as shown
in Fig. \ref{fig2}a.

It is interesting to compare the generalized Onsager method with the
Horwitz-Callen (HC) approximation which is based on the ``ring
subset'' of diagrams for the generating functional $\Phi$ in the
linked-cluster technique.\cite{HC,Wortis} In this method, the
second-order self-field $G_2$ is found by differentiating $\Phi$
with respect to the renormalized second cumulant $M_2$, while $M_2$
is represented by an integral containing $G_2$ as a parameter. This
technique does not assume any particular form for the atomic limit,
and therefore it can be used in our case including LSF as well. In
the HC method, the on-site correlator may be found as
$K_{00}=M_2+2M_2^2G_2$, and the sum rule $K_{00}=1$ is not satisfied
in the paramagnetic Heisenberg magnet. However, it is easy to check
that the value of $K_{00}$ at $T_c$ is smaller than 1 by less than a
percent in bcc and fcc lattices. In Onsager's method for the
Heisenberg model, the sum rule $K_{00}=1$ is used to fix $M_2$
instead of the integral representation as in the HC method. The
results for $T_c$ are therefore very close. We found that this close
similarity remains in the entire range of $\alpha$, as seen from
Table 1. The generalized Onsager's method is, however, technically
much simpler.

\section{Conclusions}

We have studied the thermodynamics of a simple classical spin
fluctuation model allowing for a variable degree of itinerancy. This
model is qualitatively similar to those used before to study the
thermodynamics of Fe and Ni using first-principles data.
\cite{Uhl,Rosengaard,Ruban} It is worth emphasizing that the main
drawback of using classical spin models of this type is the
ambiguity of the phase space measure. As we showed above, the
thermodynamics is very sensitive to this measure for systems with
even intermediate degree of itinerancy. While the energetics of
constrained spin configurations may, at least in principle, be
accurately mapped using DFT calculations, it is not known (to our
knowledge) how and whether the phase space measure can be supplied
in a realistic way.

In the present work, we focused on the general features of the model rather
than on the determination of its parameters from principles. We found that the
thermodynamic properties are similar to the results of the functional integral
approach. \cite{Moriya,MT78,Hubbard,Hasegawa} Further, we found that the
mean-field approximation is qualitatively valid, and short-range order is weak
and almost independent on the degree of itinerancy up to the strongly itinerant
limit where the paramagnetic susceptibility is dominated by longitudinal
fluctuations. This is in agreement with earlier results for the models of Fe
and Ni; \cite{Rosengaard,Ruban} it is clear that this is a general feature of
the classical model with no frustration.

Further, we generalized the Onsager cavity field method to itinerant systems
using an expansion around the atomic limit to first order in $1/z$. Both the
interatomic exchange constant and the Stoner parameter are renormalized by
short-range order. When both these corrections are included, the Curie
temperature is in excellent agreement with Monte Carlo results. However, simple
subtraction of the Onsager reaction field is a very poor approximation.

\acknowledgments

We are grateful to Vladimir Antropov and Nikolay Zein for useful discussions.
This work was supported by the Nebraska Research Initiative, NSF EPSCoR First,
and NSF MRSEC. K. B. is a Cottrell Scholar of Research Corporation.


\begin{thebibliography}{99}
\bibitem{Moriya} T. Moriya, \emph{Spin fluctuations in itinerant electron magnetism} (Springer, Berlin, 1985).
\bibitem{MD} K. K. Murata and S. Doniach, Phys. Rev. Lett. \textbf{29}, 285 (1972).
\bibitem{LT} G. G. Lonzarich and L. Taillefer, J. Phys. C \textbf{18}, 4339
(1985).
\bibitem{MT78} T. Moriya and Y. Takahashi, J. Phys. Soc. Japan \textbf{45}, 397
(1978).
\bibitem{Hubbard} J. Hubbard, in: \emph{Electron correlation and magnetism in narrow-band systems},
ed. by T. Moriya (Springer, Berlin, 1981), p. 29.
\bibitem{Hasegawa} H. Hasegawa, \emph{ibid.}, p. 38.
\bibitem{LL5} L. D. Landau and E. M. Lifshitz, \emph{Statistical Physics} (Pergamon, Oxford, 1980), sec. 147.
\bibitem{MW} P. Mohn and E. P. Wohlfarth, J. Phys. F \textbf{17}, 2421 (1986).
\bibitem{HE73} R. F. Hassing and D. M. Esterling, Phys. Rev. B \textbf{7}, 432
(1973).
\bibitem{Gyorffy} B. L. Gyorffy, A. J. Pindor, J. Staunton, G. M. Stocks, and H. Winter, J. Phys. F: Met. Phys. \textbf{15}, 1337 (1985).
\bibitem{Dederichs} P. H. Dederichs, S. Bl\"ugel, R. Zeller, and H. Akai, Phys. Rev.
Lett. \textbf{53}, 2512 (1984).
\bibitem{Oguchi} T. Oguchi, K. Terakura, and N. Hamada, J. Phys. F \textbf{13},
145 (1983).
\bibitem{Staunton} J. B. Staunton and B. L. Gyorffy, Phys. Rev. Lett. \textbf{69}, 371 (1992).
\bibitem{Uhl} M. Uhl and J. K\"ubler, Phys. Rev. Lett. \textbf{77}, 334 (1996).
\bibitem{Rosengaard} N. M. Rosengaard and B. Johansson, Phys. Rev. B \textbf{55}, 14975 (1997).
\bibitem{Lezaic} M. Lez\v ai\'c, P. Mavropoulos, J. Enkovaara, G. Bihlmayer, and S. Bl\"ugel, Phys. Rev. Lett.
\textbf{97}, 026404 (2006)
\bibitem{Ruban} A. V. Ruban, S. Khmelevskyi, P. Mohn, and B. Johansson, Phys. Rev. B \textbf{75}, 054402 (2007).
\bibitem{DMFT} A. Georges, G. Kotliar, W. Krauth, and M. J. Rozenberg, Rev.
Mod. Phys. \textbf{68}, 13 (1996).
\bibitem{Antropov} V. P. Antropov, Phys. Rev. B \textbf{72}, 140406(R) (2005).
\bibitem{Kubler} J. K\"ubler, J. Phys.: Condens. Matter \textbf{18}, 9795
(2006).
\bibitem{Antropov-pc} We are grateful to V. P. Antropov for his suggestion that the Anderson
criterion can be used to quantify the degree of itinerancy.
\bibitem{Binder} D. P. Landau, K. Binder, \emph{A guide to Monte Carlo simulations in Statistical Physics},
(Cambridge University Press, Cambridge, 2000).
\bibitem{Onsager} L. Onsager, J. Am. Chem. Soc. \textbf{58}, 1486
(1936).
\bibitem{Brout} R. Brout and H. Thomas, Physics (Long Island City,
N.Y.) \textbf{3}, 317 (1967).
\bibitem{Logan} D. E. Logan, Y. H. Szczech, and M. A. Tusch,
Europhys. Lett. \textbf{30}, 307 (1995).
\bibitem{Cyrot} M. Cyrot, in: \emph{Electron correlation and magnetism in narrow-band systems},
ed. by T. Moriya (Springer, Berlin, 1981); J. Magn. Magn. Mater.
\textbf{45}, 9 (1984).
\bibitem{Cyrot96} M. Cyrot and H. Kaga, Phys. Rev. Lett. \textbf{77}, 5134
(1996); H. Kaga and M. Cyrot, Phys. Rev. B \textbf{58}, 12267
(1998).
\bibitem{Wortis} M. Wortis, in \emph{Phase Transitions and Critical Phenomena},
Vol. 3, ed. by C. Domb and M. S. Green (Academic, London, 1974), p.
114.
\bibitem{VLP} V. G. Vaks, A. I. Larkin, and S. A. Pikin, Sov. Phys. -- JETP
\textbf{24}, 240 (1967) [Zh. Eksp. Teor. Fiz. \textbf{51}, 361
(1966)].
\bibitem{HC} G. Horwitz, H. B. Callen, Phys. Rev. \textbf{124}, 1757 (1961).
\bibitem{MCheisenberg} K. Chen, A. M. Ferrenberg, D. P. Landau, Phys. Rev. B \textbf{48}, 3249 (1993).

\end{thebibliography}
\end{document}